\documentstyle[preprint,aps,pra]{revtex}
\tightenlines
\def\nms{\mathsurround=0pt}
\def\overapprox#1#2{\lower 2pt\vbox{\baselineskip 0pt\lineskip - 1pt
    \ialign{$\nms#1\hfil##\hfil$\crcr#2\crcr\approx\crcr}}}
\def\gtsim{\mathrel{\mathpalette\oversim>}} 
\def\oversim#1#2{\lower 2pt\vbox{\baselineskip 0pt \lineskip 1pt
    \ialign{$\nms#1\hfil##\hfil$\crcr#2\crcr\sim\crcr}}}
\def\noal#1{\noalign{\noindent\mbox{#1}}}
\def\lambar{{\mathchar'26\mskip-9mu\lambda}}
\begin{document}
\def\etal{{\it et al.}}
\def\noal#1{\noalign{\noindent\mbox{#1}}}
\def\bq{\begin{equation}}
\def\eq{\end{equation}}
\def\bqy{\begin{eqnarray}}
\def\eqy{\end{eqnarray}}
\def\wt{\widetilde}
\def\wh{\widehat}
\title{ Zettawatt-Exawatt Lasers and Their Applications in 
Ultrastrong-Field Physics: High Energy Front}
\author{T. Tajima}
\address{ Lawrence Livermore National Lab., Univ. of California, Livermore, CA 94550\\ {\rm and}\\
 Institute for Fusion Studies, The University of Texas, Austin, TX 78712\\ \rm and\\[-10pt]}
\author{G. Mourou}
\address{ Center for Ultrafast Optical Science, University of Michigan, Ann Arbor, MI 48109}
\maketitle
\begin{abstract}
\baselineskip 24pt

Since its birth, the laser has been extraordinarily effective in the study and 
applications of laser-matter interaction at the atomic and molecular level and in the 
nonlinear optics of the bound electron. In its early life, the laser was associated with 
the physics of electron volts and of the chemical bond. Over the past 
fifteen years, 
however, we have seen a surge in our ability to produce high intensities, five to six 
orders of magnitude higher than was possible before. At these intensities, particles, 
electrons and protons, acquire kinetic energy in the mega-electron-volt range 
through interaction with intense laser fields. This opens a new age for the laser, the 
age of nonlinear relativistic optics coupling even with nuclear physics. We suggest a 
path to reach an extremely high-intensity level $10^{26-28}\,$W/cm$^2$ in the coming decade, 
much beyond the current and near future  intensity regime $10^{23}\,$W/cm$^2$, taking advantage of the 
megajoule laser facilities. Such a laser at extreme high intensity could accelerate 
particles to frontiers of high energy, tera-electron-volt and peta-electron-volt, and 
would become a tool of fundamental physics encompassing particle physics, 
gravitational physics, nonlinear field theory, ultrahigh-pressure physics, 
astrophysics, and cosmology. We focus our attention on high-energy applications 
in particular and the possibility of merged reinforcement of 
high-energy physics and ultraintense laser.
\end{abstract}

\clearpage

\section{Introduction}
\baselineskip 24pt

Over the past fifteen years, we have seen a revolution in laser 
intensities~\cite{perry94}. This revolution 
stemmed from the technique of chirped pulse amplification (CPA), combined with recent 
progress in short-pulse generation and superior-energy-storage materials like Ti:sapphire, 
Nd:glass, and Yb:glass. The success of this technique was due to its general concept, 
which fits small, university-type, tabletop-size systems as well as large existing laser chains 
built for laser fusion in national laboratories like CEA-Limeil in France, Lawrence Livermore 
National Lab. (LLNL),
Los Alamos National Lab., and Naval Research Lab. in the U.S.; Rutherford in the UK, Max Born 
Institute in Germany,
and the Institute of Laser Engineering in Osaka, Japan. A record peak power of petawatt 
($\mbox{PW}=10^{15}\,$W)
has been produced at LLNL. CPA lasers have given access to a  regime of intensities that was 
not accessible before,
opening up a fundamentally new physical  domain~\cite{mourou98}. The large leap in intensities that we have 
experienced recently is
illustrated in Fig.~1.  This figure displays the focused intensities delivered over the years 
by tabletop systems.  After
a rapid increase in the 1960s with the invention of lasers, followed by the demonstration  of 
$Q$-switching and
mode-locking, the power of lasers stagnated due to the inability to amplify  ultrashort pulses 
without causing unwanted
nonlinear effects in the optical components. This  difficulty was removed with the introduction 
of the technique of
chirped pulse amplification, which took the power of tabletop lasers from the gigawatt to the 
terawatt --- a jump of
three to  four orders of magnitude. This technique was first used with conventional laser 
amplifiers and  more recently
extended to Optical Parametric Chirped Pulse Amplifiers 
(OPCPA)~\cite{dubeis92}.  A number of laboratories 
are presently
equipped with CPA ultrashort-pulsed terawatt lasers such as Laboratoire Optique Applique 
in France, University of
Lund in Sweden, Max-Planck Institute in Garching, Jena University, and the Japan Atomic 
Energy Research Institute
in Kansai. The CPA-enabled short-pulse generation has advanced to the single-cycle 
regime~\cite{morgner99}. The peak power in
the 1990s reached 100~TW, demonstrated at JAERI Kansai.  More recently, deformable
mirrors have been
incorporated into CPA,  making it possible with low f\# parabola to focus the 
laser power on a
$1~\mu\,$m spot size~\cite{druon98}.  Present systems deliver focused intensities in the
$10^{20}\,$W/cm$^2$ range. In the near future,  CPA systems will be able to produce 
intensities of the
order of $10^{22}\,$W/cm$^2$. As indicated in  Fig.~1, we will see a leveling off of
laser intensity for
tabletop-size systems at $10^{23}\,$W/cm$^2$. This limit~\cite{perry94} is imposed, 
as we will explain later, by the
saturation fluence - energy per unit  area --- of the amplifying medium and the damage
threshold of the optical
elements. In a well-conceived CPA system the saturation fluence is of 
the order but less than the damage threshold. Once this limit is reached, we will have
accomplished a leap in intensities of eight orders of
magnitude and the only way to increase the focused intensity  further will be by 
increasing the beam size,
leaving the following questions to be answered:  How could we go higher in intensity?
And how much higher?

Because the highest intensities will rely on the largest pump available, we explore 
if it could be technically
feasible to build a large scale CPA (OPCPA) pumped by a megajoule system of the type
of the NIF (National
Ignition Facility) in the U.S. and the LMJ (Laser Megajoule) in France. Power in the 
zetta $(10^{21})$ watt range  could be produced, yielding a focused intensity of 
$10^{28}\,$W/cm$^2$. These
intensities well beyond the current intensity accessible will open up a 
new physical regime.

\section{ The theoretical intensity limit}

In a CPA system the maximum energy per pulse obtainable is limited (1)~by the damage 
threshold $F_d$ of the stretched pulse --- of the order of a few nanoseconds, and (2)~by the 
saturation fluence $F_{\rm sat}= h\nu/\sigma$ of the amplifier. Here $h$ is the Planck constant,  
$\nu$ is the
laser  frequency, and $\sigma$ the cross-section of the lasing transition at the lasing frequency 
$\nu$, $F_{\rm
sat}$ corresponds to the energy per unit area, necessary to decrease the population inversion by a 
factor of two, in a time shorter than the excited state lifetime, $T_f$. For stretched pulses of the 
order of a nanosecond, $F_d$ is of the order of $20\,$J/cm$^2$ and $F_{\rm sat}=1\,$J/cm$^2$ for 
Ti:sapphire.
For  Yb:glass, $F_{\rm sat}=40\,$J/cm$^2$ --- significantly greater than the damage threshold 
of most optical 
components. In that regard, high-damage-threshold materials need to be developed in order to 
take full advantage of the excellent energy storage capability of this material. The minimum 
pulse duration $\tau_p$ is imposed by the Heisenberg uncertainty relation $\Delta\nu_a 
\tau_p=1/\pi$, where
$\Delta\nu_a$ is  the transition bandwidth for a homogeneously broadened bandwidth. 
The maximum power per 
unit area is, therefore, given by
\begin{equation} 
P_{th}=\pi\,\frac{h\nu}{\sigma}\,\Delta\nu_a.
\end{equation}
While the maximum focusable intensity $I_{th}$ will be obtained when this power is focused on a 
spot size limited by the laser wavelength $\lambda$, leading to the expression
\begin{equation} 
I_{th}=\pi\,\frac{h\nu^3}{\sigma}\cdot\frac{\Delta\nu_a}{c^2}.
\end{equation}
This intensity limit is represented in Fig.~1.

\subsection{The zettawatt system}

Although a zettawatt system could be built using Yb:glass, with the advantages of being 
relatively compact due to the high $F_{\rm sat}$ of this material and being diode pumpable, much 
development work needs to be accomplished to reach this intensity level with this material.
The proposed systems described below have been stimulated by the construction , both in France and 
in the
U.S,  of lasers delivering a few megajoules of energy as well as the availability of large telescope 
technology (10m diameter) and deformable mirrors.

Let us recall that the NIF and LMJ systems will deliver 2~MJ at 350~nm, i.e., the third harmonic 
of 1060~nm in 3~ns. The energy at 530~nm, for a long pulse of 10-20~ns, could be as high as 5 
MJ. This energy could be used to pump a CPA-type system to produce of the order of 1 MJ of 
energy in 10~fs on a spot size of $1\,\mu\,$m~\cite{druon98}. Such a system will have a power of the order 
$10^{20}\,$W or 
100 exawatts with an intensity of $10^{28}\,$W/cm$^2$. To produce these phenomenal characteristics, 
we have two alternatives. The first one would be to use this large pump energy to drive an 
OPCPA system. This elegant technique~\cite{dubeis92} has been demonstrated to the joule level at 800~fs
level a few
years ago by I. Ross \emph{et al.} from Rutherford~\cite{ross97} and is being implemented in the same
laboratory to the 10~PW,  30~fs level. It has the advantage that it could use KDP
as the nonlinear 
medium. KDP has been produced in large dimensions at a relatively low cost for the NIF and
LMJ programs.
The working fluence on the crystal will be 1~J/cm$^2$, leading to a beam diameter  of 
around $10\,$m$^2$
to accommodate the whole pump energy. This technique has the potential to  produce a 
10-fs pulse or shorter,
but needs to be demonstrated at least at the joule level. The  second approach, though
more conservative, will
be to drive a CPA Ti:sapphire, a well-established technology. The amplifier will be composed of a large matrix
of Ti:sapphire rods. The size of the matrix will be dictated by the saturation fluence $F_{\rm
sat}=1\,$J/cm$^2$, corresponding  to a $10\,$m diameter beam. The amplifier could be composed of 2500,
$20\times 20\,$cm pieces, each $2\,$cm in length.  Each piece will have to be segmented in order to avoid
the transverse amplified spontaneous emission. Ti:sapphire of
$20\times 20\,$cm has already been grown~\cite{felt}. In both schemes the beams  can be focused by a large parabola. The
dielectric coating used for the parabola will have a  damage threshold 
of 1~J/cm$^2$ (for short pulses)~\cite{du94,stuart95},
imposing a parabola size of $10\,$m in diameter. This parabola will have the same diameter as the Keck telescope. The
phase  front will be interferometrically controlled by an active matrix of deformable mirrors. The  grating compressor
could be made out of meter-size gratings assembled interferometrically in a  matrix geometry. The size of the grating
for a megajoule short-pulse system will be also of the  order of $10\,$m in diameter, dictated by a damage threshold of
1~J/cm$^2$ (such a damage threshold  has just been 
reported~\cite{migus}). Each large grating will be composed of
100 of 1~m$^2$-size gratings. 

Note that although large, the number of optical components involved will be small, compared 
to a MJ system calling for 4500 laser slabs, 800 large KDP frequency converter crystals, and 
500 gratings of meter-size. An alternative method using a plasma has been 
suggested~\cite{felt}, but will lead to the same
output energy limited by the pump. In addition the beam must go through a plasma, which is highly undesirable.

\subsection{ The exawatt laser system}

If a zettawatt laser, although feasible, could seem too grandiose at this time, an exawatt system on the 
other hand,which would produce $10\,$kJ in $10\,$fs, i.e., $10^{25}\,$W/cm$^2$, could be readily constructed. 
Only one percent or $30\,$kJ of the NIF/LMJ energy would be necessary. The beam size will be 
of the order of one meter in diameter. The amplifying method will be composed of a matrix of 
25 Ti:sapphire $20\times 20\,$cm$^2$ crystals~\cite{felt} and two gratings of meter-size. The segmented telescope will 
have a one-meter aperture. The wave front will be corrected by a large deformable mirror. 

In the following we explore a few examples of applications of such intense lasers that may enable new ways to
investigate fundamental physics.

\section{Fundamental Physics: Applications of Zettawatt and Exawatt Lasers}

The major signpost of contemporary high-field science is the entry into the relativistic regime, 
which is characterized by the quivering momentum of electrons in the laser fields reaching the 
speed of light times the electron rest mass. This field is reached when the laser intensity is on 
the order of $10^{18}\,$W/cm$^2$ for a typical optical frequency of the laser we mentioned above. When 
the laser field is much less than this, free electrons behave harmonically to the field's optical 
oscillations. Although these oscillations of electrons are still important enough to couple with 
various collective motions of free electrons, there is small orbital 
nonlinearity associated with the quivering motion in this regime. For bound electrons in atoms 
or molecules the quivering motion in the laser fields is once again, by and large, a perturbation 
on their orbital motion. This perturbation is enough to cause various interesting phenomena in 
the non-relativistic regime. There is such a wealth of phenomena known in this regime that it is 
hard to list them briefly, but they include the multi-photon process of ionization, that of 
transitions, Raman and Brillouin scatterings, and various optical 
nonlinearities~\cite{bloem65} arising from 
the material's response. In the relativistic regime with the intensity above $10^{18}\,$W/cm$^2$, in 
addition to the above phenomena, there emerge new classes of effects largely arising from the 
relativistic nonlinearities of electrons in the high field.

The field intensity in this regime means that the electron momentum in the light is 
typically $e E_0\omega_0$, which becomes of the order of and exceeds $m_0c$, where
$\omega_0$ is the laser angular  frequency, $E_0$ the laser electric field, and $m_0$ the electron 
rest mass. When the electron  momentum exceeds $m_0c$, the electronic orbit ceases to be harmonic and linear. It
becomes a  figure-8 motion, including higher-harmonic components. The photon pressure is exerted 
individually on an electron by the electron-photon collision through the Thomson cross-section
\begin{equation} 
\sigma_{T}={8\pi\over 3}\cdot{e^4\over m_0^2 c^4}\approx 7\cdot 10^{-25}\,\mbox{cm}^2.
\end{equation}
When the flux of laser at the intensity entering the relativistic regime 
is shone on an electron (with Lorentz factor $\gamma$), 
this causes a force on it
\begin{equation} 
F={\sigma_{T}\over\gamma}\,{E^2_0\over 4\pi}.
\end{equation}
When the laser intensity is $I=10^{26}\,$W/cm$^2$, the force acting on an electron is $10^{-2}\sim 10^{-1}\,$erg/cm. 
Or the acceleration acting on an electron (originally) at rest is
\begin{equation} 
a_e={f\over m_0}\sim 10^{25-26}\,\mbox{cm/s}^2,
\end{equation}
where $m_0$ is the rest mass. [In contrast to this, the Schwinger acceleration is 
\begin{equation} 
a_S = 2\cdot 10^{31}\,\mbox{cm/s}^2,
\end{equation}
at which an electron gains energy by $m_0c^2$ over the Compton length $\lambda_C=\hbar/m_0c$, and the pair
creation becomes prevalent.] A large flux of photons bombarding an electron causes such high acceleration through 
collisions between photons and electrons (Compton collisions). Such acceleration may be 
called the Eddington acceleration, as Eddington introduced the stellar luminosity at which the 
gravitational pull is balanced by this photon collisional acceleration (the Eddington 
luminosity). 

The Eddington acceleration may be likened to the acceleration of water molecules near the 
surface of water in a lake when there is a breeze passing over the water surface. The collisional 
viscosity created by collisions between water molecules and wind molecules gives rise to a 
water flow. However, we also observe that when the breeze gets stronger or becomes a gale, 
the water surface is no longer smooth and acquires ripples or waves. Such ripples facilitate an 
increase in the effective viscosity of water for the wind, so that the wind momentum is 
anomalously effectively transported to water molecules with a much faster rate through the Kelvin-
Helmholtz instability at the interface of the two fluids. What happens in the case of strong 
photon flux (``wind") in a plasma? Just like the strong wind on the water surface, the strong 
photon flux is capable of creating plasma waves, which in turn causes enhanced viscosity and 
thus anomalous momentum transport from photons to electrons of the plasma. This latter 
process is through a collective interaction. The plasma wakefield 
excitation~\cite{tajima79} is typical of 
this, in which plasma waves generated in this process are accentuated, and new processes of 
collective interaction emerge. In this ultraintense regime, electrons 
may be accelerated not only through the electrostatic field that is 
set up by the ponderomotive force of the laser, but also directly by 
the ponderomotive force itself to very high energies. If we apply 
this laser at the resonance absorption at high densities near the 
compressed laser fusion target, on the other hand, much of the laser 
energy may be converted into relatively low-energy copious electrons, 
which could constitute a new alternative to the fast ignition fusion. 
The acceleration of heavier particles (protons and other nuclei) to 
relativistic energies will become possible, too. Either by direct 
baryon acceleration by this, or other process, (it will take a variety of 
experimental realizations such as a target irradiation, cluster 
irradiation, converging imploding shells, etc), we will access the 
nuclear regime of matter reminiscent of the early epoch of the Big 
Bang (see Fig.~1). The production of extremely high energy or copious 
gamma rays will happened. The extreme high photon pressure (which already exceeds 
Gbar in the presently available intense lasers) may be 
finally directly utilized to directly compress matter in this 
ultrarelativistic regime, because baryons, too, become relativistic. If 
so, unprecedented densities of matter may be created. The combination of extreme intensity lasers and 
high-energy particle beams that can be created by the conventional 
high-energy physics accelerator will further multiply our ability to 
expand our frontier horizon. This will be the merging point of 
high-energy physics and high-field science. In the following, we list several examples of exciting new 
frontiers of fundamental physics that may be explored by this regime of 
intensities. In many of the applications we discuss, in this 
high-intensity regime the interaction length between the laser and 
matter is expected to be extended beyond the Rayleigh length, as the 
relativistic mass effect of electrons sets the self-focusing threshold 
at $10^{10}(\omega/\omega_{p})^{2}\,$W~\cite{sprangle87}.

\subsection{ Particle acceleration}

The pulse or self-modulation of a photon wavepacket with sufficient intensity induces a 
longitudinal electric field (in the $x$-direction, as part of plasma oscillations) as 
\begin{equation} 
E_x\sim\sqrt{\frac{n_e}{n_{18}}}\ 
a^2_0\qquad\mbox{(GeV/cm)},\qquad\quad (a_0 \le 1),
\end{equation}
where $n_{18}=10^{18}$/cm$^3$ and $n_e$ is the electron density, $a_0=eE_0/m_0\omega_0 c$   the
normalized vector potential  of the laser, which is sometimes called the quivering velocity normalized to $c$
(or quivering  momentum normalized to $m_0c$). The energy gain over the interaction length $l_x$ is
\begin{equation} 
\Delta\varepsilon\sim\sqrt{\frac{n_e}{n_{18}}}a^2_0\ell_x\,\mbox{(GeV)}.
\end{equation}
When $a_0$ exceeds unity (ultrarelativistic; $I\ge 10^{18}\,$W/cm$^2$), the pressure of the photon 
wavepacket becomes so large that nearly all electrons are evacuated from 
the laser packet~\cite{ashour81}. The  photons
plow through the plasma with electrons piling up in front of the pulse (``snow plow" of electrons), yielding the 
snow plow acceleration~\cite{tajima85} with momentum gain of electrons
\begin{equation} 
p_x={1\over 4}\,{E^2_0\over n_e c}={E_c^2\over n_e m_0 c^2}\,a^2_0,
\end{equation}
where $E_c=m_0\omega_0c/e$. Here the laser pulse is assumed to have fully interacted with the plasma. If 
the pulse quickly diffracts before the full interaction, say over the Rayleigh length, $p_x$ is simply 
proportional to $E_0$.

Now let us imagine, for the moment, that the laser has only half the period 
(unipolar)~\cite{scheid89} or subcyclic~\cite{rau97}. In this case the energy (or momentum) gain is 
\begin{equation} 
\Delta\varepsilon\sim m_0c^2 a_0^2. 
\end{equation}
Since the Lawson-Woodward theorem~\cite{lawson79} prohibits any overall acceleration for fully oscillatory (i.e., 
usual) electromagnetic waves in vacuum in infinite space, the above energy gain is 
compensated for by decelerating phase. There are, however, many instances that break the theorem requirements.
For example, it may be possible in this extreme relativistic  regime that electrons are accelerated to very high energy,
immediately reaching the speed of  light and becoming in phase with the photon over a sufficiently long distance, so
that by the time  they become dephased, the EM wave may decay away for some reason, such as by radiative 
decay or pump depletion to the acceleration. If this happens, the above energy gain (or a 
portion of it) may be preserved. In extreme high-intensity regimes such effects will become significant.  If
the laser spreads  over the Rayleigh length, the energy gain is proportional to $a_0$. The transverse
momentum gain $(p_y)$ is always proportional to $a_0$. In a broad 
general way we can say that the interaction between the laser and 
electrons becomes more coherent, as the laser intensity increases 
because the laser and electrons move more coherently over greater 
distance. This is the signature of the acceleration by the 
ponderomotive potential of photon fields in ultra-relativistic 
intensity regimes.

Thus it is possible to see electrons at energies of up to $\sim
100\,$TeV at the laser intensity of  $10^{26}\,$W/cm$^2$ and even up to $\sim 10\,$PeV at
$10^{28}\,$W/cm$^2$. The  accelerating gradient is $200\,$TeV/cm and $2\,$PeV/cm, respectively. 
Note that
such energies (100~TeV and 10~PeV) if collided, correspond to $10^{19}\,$eV and $10^{23}\,$eV 
for fixed target
experiments. These energies rival or exceed those of the highest energy cosmic rays, which 
are observed up to
$3\times 10^{20}\,$eV. Of course, it is not easy to attain correspondingly high luminosity for collisions at such high
energies. Even though the exploration of particle physics at pb may not be within reach, we may use such particles in
pursuit of (other) fundamental physics at the energy frontier. We might recall that Anderson first discovered mesons
in cosmic rays, followed by more detailed studies of those particles in cyclotrons and other accelerators. Perhaps the
present way-out parameters in the energy frontier may herald some new phenomena. One such example may be the test
of Lorentz invariance~\cite{sato72} in extreme high energies. For such a test, 
unlike the detection of new particles with pb cross-section, the 
luminosity requirement may be much relaxed.

When we irradiate an extremely relativistic laser pulse $(I\sim 10^{26}\,$W/cm$^2$) on a thin film
$(\gtsim 1\,\mu\,$m) of a metal at a tight spot of ($1\, \mu\,$m)$^2$ followed by a microhole in a
metallic slab over more  than one cm, we hypothesize that the laser pulse picks up metallic electrons from the
film and  continues to propagate through the microhole, as it remains focused. If this proves to be the  case,
the amount of electrons that are to be accelerated by this pulse is in the ballpark of 
\begin{equation} 
N_e\sim n_eAl_b\sim 3\cdot 10^{10},
\end{equation}
where we assumed $n_e\sim 10^{24}\,$cm$^{-3}$, area $A=(1\, \mu\,$m)$^2$, the bunch length 
$l_b\sim 3\cdot
10^{-6}\,$cm. In such a large  pickup, the pump depletion due to the energy transfer to electrons 
is significant. 
In fact, it  may play a fundamental role in this acceleration to turn the electromagnetic energy into 
particle kinetic energy without returning to the decelerating phase, providing one way to break 
the Lawson-Woodward theorem's constraint, as we mentioned earlier. If we take 1/10 of the 
above electrons to gain
10  TeV and if we focus electrons (and positrons) down to $10^{-6}\,$cm (or $10^{-7}\,$cm) 
at focus (the collision 
point), the luminosity of the colliding events is of the order of
\begin{equation} 
{\cal L}=10^{31}\,f\ /\mbox{cm}^2/\mbox{s}, (\mbox{or}\ 10^{33}f),
\end{equation}
where $f$ is the collision repetition rate. Nakajima has considered a similar 
but more daring luminosity scenario~\cite{nakajima01}.

\subsection{Fast Ignition Fusion}

One special case of laser energy conversion into electrons is through 
the resonance absorption at the critical density. The concept of fast 
ignition in laser-driven inertial fusion~\cite{tabak94} calls for 
laser beam of $\sim 10\,$psec duration at the intensity exceeding 
$10^{20}\,$W/cm$^{2}$ to be absorbed at the critical density $(\sim 
10^{21}-10^{22}$/cm$^{3})$, creating a beam of electrons in the 
several MeV range. The idea is to separate the roles of laser into 
two functions: one to compress the fuel with least amount of entropy 
increase so that the fusion fuel is compressed to a highest density 
with least amount of laser energy, and the other is to heat the fuel 
to the thermonuclear ignition temperature ($\sim 10\,$KeV) when the main 
compression is achieved. The latter  step may be carried out 
according to Tabak \etal\ by the appropriate range of energetic 
(several MeV) electrons that are transported from the crust of the 
target $(<10^{22}$/cm$^{3})$ to the surface of the fuel (at $\sim 
10^{26}$/cm$^{3}$) at the pulse duration of $\sim 10\,$ps. In order 
for electrons to trigger the fusion ignition, the condition
\bq
\rho R\alt 0.5\qquad\left({\rm g\over \rm cm^{2}}\right)
\eq
has to be 
fulfilled~\cite{tabak94}. Here $\rho$ is the density of the compressed 
fuel and $R$ the electron range and thus approximately the size of the 
fuel at compression. Although this idea is potentially capable of 
reducing the necessary laser energy by almost an order of magnitude 
or increasing the fusion gain by an order of magnitude at the same 
laser energy, there remains a considerable uncertainty in the efficiency of 
energy conversion from short-pulse (10ps) laser to electron beam to the 
compressed fuel and in the stability and reliability of electron beam. 
For example, laser and electron beam have to propagate 
through the over dense plasma, through which filamentation and 
kinking instabilities are found to arise. In order to cope with this 
problem, the ignitor laser is further split into the hole boring one 
and energy deliverer.

We suggest that an alternative method of fast ignition by much 
shorter-pulse laser may be possible. In this, although the total amount 
of energy necessary to be delivered is unchanged ($50\sim 100\,$kJ), we 
shorten the pulse length to 10fs so that the local laser intensity 
reaches of the order of $10^{25}\,$W/cm$^{2}$. Since the resonance 
frequency reduces inversely proportional to $\sqrt{n_{e}}$, the 
resonance density becomes in the order of $10^{25}$/cm$^{3}$, a very 
close proximity of the fully compressed fuel. This way we may be 
avoiding the difficult and long energy transport of electron beam from 
the density region of $<10^{22}$/cm$^{2}$ to $10^{26}$/cm$^{3}$. With 
the recent success of improved target design with a conic aperture for 
fast ignitor laser beam access~\cite{kodama01}, our ultrafast approach 
may be further bolstered as a direct energy delivery vehicle. It 
remains to be seen, however, how much fraction of the laser energy is 
consumed due to the pump depletion~\cite{horton86} while it interacts 
with the surrounding plasma. It also needs to be investigated what 
kind of electron energy spectrum is generated in the ultra-intense 
laser beam. The mission of the electron energy conversion in fast 
ignition is orthogonal to the previous Subsec.~3A of super-high energy 
electron generation. The production of a small fraction of extremely 
high-energy electrons is tolerated, as long as the majority of energy 
is in several MeV electrons.

\subsection{ Baryon acceleration}

Many thought it difficult to accelerate protons and heavier particles by light, as massless light 
propagates at the speed of light, while protons are massive and nonrelativistic --- until last year, 
when the Petawatt Laser experiment~\cite{key99} and other 
experiments~\cite{maksim00,clark00,snav00} showed that 
protons have been accelerated much beyond a mega-electron-volt.  The 
observed transverse emittance is about 0.5~mm~mrad, while the 
longitudinal one is about  MeV-psec~\cite{roth01}. These early 
experiments already rival or even surpass those of the 
conventional ion sources in some of crucial parameters. The main mechanism of 
laser proton acceleration in the above experiments is due to the space charge set up by 
energetic electrons that are driven forward away from the back surface of the target slab. The 
energy of protons is thus dependent on that of electrons. Bulanov and others 
showed~\cite{esirk99} in 
simulation that at a laser intensity of $I=10^{23}\,$W/cm$^2$, protons are accelerated beyond a
giga-electron-volt. If this process of proton acceleration scales with the  intensity (as the
electron energy does), we may be able to see 100~GeV protons and 10~TeV at $I=10^{26}$ and
$10^{28}\,$W/cm$^2$, respectively. However, it may also be possible that this process is now 
due directly to the
photon pressure beyond the intensity regime of
$I=10^{24}\,$W/cm$^2$. The energy  expected through this mechanism is about the same as that through the space
charge  mechanism.

It is not clear how much energy will be in protons. In the Petawatt Laser experiment, about 
10\% of laser energy (300~J) was converted into proton energy --- 30~J 
(beyond 1~MeV)~\cite{key99}. If 
we take this conversion efficiency in the extreme relativistic laser intensity, then more than 1~kJ of proton energy is
expected for the case of intensity $10^{26}\,$W/cm$^2$. If we further take a flat  energy spectrum, approximately
$10^{11}$ protons are accelerated beyond 10~GeV in this intensity  
regime. If we can generate the solitary accelerating 
structure, such energy may exceed 100~GeV. If we can 
converge these in a colliding pair of beams at focus, the luminosity 
of colliding hadrons is ${\cal L}=10^{34}\,f$, if we focus on 10~nm. 
The expected number of nuclear events is on the order of $10^{9}$ per shot. 

Such an intense, relativistic, compact proton source has a number of fascinating applications. It 
may be applicable to ion radiography, fast ignition of fusion, etc., among many others. An 
additional application is pion (or muon) and neutrino beam generation.  With sufficiently 
relativistic proton energies the
emittance of created pions can be sufficiently small. If so, they may be 
promptly accelerated pions to sufficiently high
energies before the space charge effect expands the beam emittance and 
before they die out. In this the prompt
acceleration and its compactness are important, both of which are the 
forte of the laser accelerations in an application. Protons beyond a 
certain energy (several hundred MeV) in matter induce through the 
nuclear strong interaction the creation of pions, which in turn decay 
into muons and neutrinos in a matter of 20~ns, if nonrelativistic,
propagating mere 6~m at most:  $\pi\to\mu+\nu$. With the conventional 
accelerating gradient of, say, 20MeV/m over this distance of 6m, we 
can increase the pion energy by 120MeV, which will increase the 
lifetime of pions but not by an order of magnitude. On the other 
hand, the laser acceleration with its far-greater gradient would 
increase the energy and lifetime of pions far more than this (and 
also reduce the emission cone angle). This would contribute to a 
further smaller emittance, which in turn contributes to higher energy, 
lower emittance muons and neutrinos. 

\subsection{ Gamma ray emission}

The emission of gamma rays from the intense laser is expected. Although the well-known 
bremsstrahlung x-rays (and gamma rays) by electrons through the collision with nuclei are 
expected to remain important, the Larmor radiation is the most intense in the extreme 
relativistic regime among all radiation mechanisms through the 
interaction with matter (in this case free electrons) with power
\begin{equation} 
P_L=\left({8\over 3}\right)r_e m_0 c\omega_0^2 a_0^2.
\end{equation}
Gamma rays (whose energy peaks at $\hbar\omega_{L}\sim 
a^{3}_{0}\hbar\omega_{0}$) may be forward directed as electrons are accelerated forward. In addition, gamma 
rays of nuclear origin are also expected. As discussed in Subsec.~3A, 
accelerated electrons with extremely large Lorentz factor created by 
laser may be easily converted into extreme high-energy gamma rays 
(polarization may be preserved if desired). 

When an intense laser is directed at a high-energy electron beam, the 
well-known energy enhancement of laser photon happens by the factor 
$\gamma^{2}$  or up to the electron energy itself (where $\gamma$ is the Lorentz factor of the head-on 
colliding electron beam) through Compton scattering. 
Utilizing this, high-energy electron (or positron) colliders may be 
converted into a $\gamma$-$\gamma$ collider~\cite{coleman99}. The idea of a 
$\gamma$-$\gamma$ collider is an example of the cooperative employment of two 
separate technologies (the indigeneous high-energy physics and the 
high-field science) to enhance the overall goal.

Gamma rays in the $\gamma$-$\gamma$ collider mode or with an enormous energy 
as discussed in Subsec.~3A (with another laser, for example) may be employed to test the Lorentz 
invariance~\cite{sato72,telnov82}.  For example, if Lorentz invariance is 
violated, the event number of $e^{-}e^{+}$ pair creation is suppressed.

\subsection{Superhot matter}

Kishimoto and Tajima~\cite{kish00} have shown that an intense laser pulse may be nearly totally 
absorbed by only several layers of atomic clusters. When the excursion length of electrons 
exceeds the size of clusters, electron orbits become chaotic upon their removal from the 
original site of the cluster, due to the cluster polarization. The chaos sets in within a few 
femtoseconds, thus making the absorption of the laser ultrafast. Further, upon removal of 
electrons, ions of the cluster Coulomb-explode, gaining a large fraction of electron energy. If 
we arrange matter in such a way as to absorb nearly all laser energy over the thickness of a few 
microns on a $(1\,\mu\,$m)$^2$ spot, the average energy per particle is approximately 100~GeV and $10^4\,$GeV,
at $I=10^{26}$ and $10^{28}\,$W/cm$^2$, respectively. Such superhot matter is expected to generate  copious
positrons through the Breit-Wheeler process and perhaps other nonlinear quantum  electrodynamic (QED)
processes~\cite{taguchi}. Probably the amount of positrons generated will not be less  than $10^9$~(see also in
Subsec.~3F).

A large number of nuclear excitations are expected. In fact, in the Petawatt 
Laser experiment~\cite{cowan00} nuclear
transmutations through generated gamma rays have been observed. Since the  electron energy is much greater than
the typical nuclear excitation energy, copious reactions  are expected. The nuclear reactivity $Y_e$ is
estimated~\cite{udagawa96} as
\begin{equation} 
Y_e=1.5\cdot 10^{12}\,\frac{n_i n_e \sigma\tau_p\ell}{n_{i0} n_{e0}\sigma_0\tau_0\ell_0}\,\eta.
\end{equation}
Here $n_i$ and $n_e$ are the ion and electron densities, $\sigma$, the reaction cross-section, $\tau_p$ and
$\ell$ are the interaction time and length, and  $\eta$ is the fraction of reaction electrons.  The numbers in the
denominator  are the normalizing numbers with typical parameters as follows: $n_{i0}=5\cdot 10^{22}$/cm$^3$, 
$n_{e0}=10^{24}$, $\sigma_0=1\,\mu\,$b,  $\tau_0=10^{-12}\,$s,  $\ell_0=10^{-3}$. If we take $\ell\sim 1\,$cm, 
$\eta= 0.5$, $\sigma\sim 1\,$mb, we obtain $10^{14} >Y_e>10^{10}$. For example, the element of Hf is excited at
energy around 1~MeV. 

Even if a heavy metal is irradiated, the energy per nucleon exceeds 1~GeV. Thus by irradiating 
the target from both ends, it may be possible to cause colliding heavy nuclei. The possible 
nuclear events we expect are
\begin{equation} 
N_n=n_i^2 c\sigma V\tau_p\sim 10^7
\end{equation}
where  $V$ is the interaction volume taken as $10^{-8}\,$cm$^3$. We expect about $10^7$ events per laser shot
of nuclear events, which may include such a process as quark-gluon  plasma formation. The presumed transition
temperature from nucleon state to  deconfined quark-gluon plasma state is $\sim 200\,$MeV. The relativistic heavy ion
collider (RHIC)  delivers $10^9$ ion beams (Au) at luminosity of $2\cdot 10^{26}$/cm$^2$/s, yielding a few events of
quark-gluon  plasma per second.

\subsection{ Super-high pressure}

The photon pressure at the extreme relativistic regime is immense: 30~Pbars ($= 3\cdot 10^{10}\,$Mbars) 
and 3~Ebars, at $10^{26}$ and $10^{28}\,$W/cm$^2$, respectively. With such extraordinarily high pressure, one 
may be able to access entirely new regimes of highly-pressurized matter states. The current 
method of achieving the highest pressure and compressed states is to ablate exterior matter 
outward by laser in order to drive interior matter inward by the reaction momentum. This way, 
gradual (nearly) isentropic compression to a density $10^3$ higher than the solid density can be 
achieved with the laser whose pulse is longer than a nanosecond. The laser light, as well as hot 
electrons, is absorbed before the interior matter; the formation of a strong shock, which is 
detrimental to compression, is avoided. However, the superstrong pressure in the extreme high-field regime, as well
as  another set of parameters that come with it, invites us to think about and explore radically different ways 
than the ablative method to create ultrahigh compressed states. This 
may allow us to access densities far greater than $10^{26}$/cm$^{3}$.

\subsection{ Nonlinear QED and horizon physics}

At the intensity $10^{28}\,$W/cm$^2$, the electric field is only an order of magnitude less than the 
Schwinger field as discussed below. At this field, fluctuations in 
vacuum are polarized by laser to yield copious pairs of real electron and positron. In collider 
physics a similar phenomenon happens when the so-called $\Upsilon$ parameter reaches unity. In 
reality, even below the Schwinger field, the exponential tail of these fluctuations begins to 
cause copious pair productions.

Though pair production has been demonstrated at SLAC 
(E144-experiment)~\cite{bula96} by the interaction of $\gamma$-ray
with an intense laser at  intensities in the range of $10^{18}\,$W/cm$^2$, the direct production 
of pairs by a high-intensity laser from vacuum remains elusive. The rule of thumb for threshold of  pair production derives from the
simple argument that it is the field necessary for a virtual electron to gain an energy $2m_0 c^2$ during its lifetime
$\delta t$, imposed by the Heisenberg uncertainty principle $\delta t=\hbar/m_0 c^2$, the
energy gain length,  $c\delta t$ is the Compton length $\lambar_c$. Hence, the breakdown field
$E_S$, the  Schwinger field, is $E_S=m_0 c^2/e\lambar_c$  where
$\lambar_c=0.386\,$pm $E_S=2\cdot 10^{16}\,$V/cm. (The laser field $E_S$ is related to the laser
intensity $I_\ell$ by $E^2_S=Z_0 I_S$ where $Z_0$ is the vacuum impedance. 
For  $Z_0=377\,\Omega$, we find
a value of $I_S=10^{30}\,$W/cm$^2$).

This approach gives an estimate for the threshold for pair creation and do not provide the number of pairs that could be
created for a given intensity. The probability of spontaneous production of pair creation per unit time per unit volume
by Schwinger~\cite{schwinger51} is
\begin{equation}
w={1\over\pi^2}\,{\alpha\over\delta t}\,{1\over\lambar^3_c}\left({E\over E_S}\right)^2\sum^\infty_{n=1}\,{1\over
n^2}\,\exp\left(-n\pi\,{E_S\over E}\right),
\end{equation}
where $\alpha$ is the fine structure constant. The number of pairs $N$ for a given laser field is
\begin{equation} 
N=V\tau_p w
\end{equation}
where $V$ is the focal volume. For $V=10^{-12}\,$cm$^3$ and $\tau_p=10\,$fs,
we find the generation  of $10^{24}$ pairs at the Schwinger intensity $I_S$. It is interesting to note that the
intensity to create a single pair is still the gargantuan intensity of $10^{27}\,$W/cm$^2$. However, as indicated
earlier (in Subsec.~3E), the presence of matter such as nuclei or electrons reduces this field by a considerable
amount. 

The interaction of intense laser with high-energy electrons will 
enhance some of the parameters even further. For example, 
counterstreaming electron beam and laser can produce copious 
polarized (high-quality) positrons.  This can serve as a polarized 
positron source for one thing. In an extreme field regime, on the 
other hand, the laser field is enhanced 
by the Lorentz factor of the electron beam, so that the effective 
field from the electron frame may far exceed the Schwinger value. If 
this field is exceeded by much more than these orders of magnitude, 
direct production of other particles such as muons out of  
``vacuum'' may be observed.

In addition to the test of nonlinear fields, we are able to explore what may be called `horizon physics.' According to
Einstein's equivalence principle, a particle that is accelerated feels gravity in the  opposite direction of the
acceleration. The acceleration due to the electric field of the laser at  this intensity is huge: $a_e\sim 10^{30}$ and
$10^{31}\,$cm/s$^2$, at $I=10^{26}$ and $10^{28}\,$W/cm$^2$, respectively. An  observer at rest (or in an inertial
frame of reference) sees the horizon at infinity if there is no  gravitation. On the other hand, an observer near a black
hole sees the horizon at a finite  distance where the gravitation diverges. Equivalently, an observer who is being
accelerated  (feeling immense equivalent gravity) now also sees the horizon at a finite distance. Any 
particle (``observer" --- a wave function) that has a finite extent has one side of its wave function 
leaking out of the horizon. The Unruh radiation is emitted when this 
happens~\cite{unruh76}. Unruh 
radiation is a sister to the Hawking radiation~\cite{hawking74}. The Unruh temperature
\begin{equation} 
k T_U={\hbar a_e\over 2\pi c}
\end{equation}
which is about $10^4\,$eV and $10^5\,$eV, for $I=10^{26}$ and $10^{28}\,$W/cm$^2$, respectively.  The
radiative  power of Unruh radiation increases in proportion to $a_0^3$ (or $I^{3/2}$), in contrast to the Larmor 
radiation power of $a_0^2$~\cite{chen99}
\begin{equation} 
P_U={12\over\pi}\, {r_e\hbar\over c}\, a_0^3\omega^2_0,
\end{equation}
where $r_e$ is the electron classical radius. Since in the extreme relativistic regime the radiation is 
dominated by Larmor radiation, the Unruh signal has to compete with the Larmor with the ratio  of powers
\begin{equation} 
{P_U\over P_L}={\hbar\omega_0\over m_0 c^2}\, a_0.
\end{equation}
At this regime, the Unruh is down only by a few orders of magnitude, and it has been 
suggested to circumvent the noise to observe the Unruh signal by exploiting polarization, 
etc.~\cite{chen99}. This $P_U$ yields $10^4\,$eV/s and $10^7\,$eV/s, for $I=10^{26}$ and $10^{28}\,$W/cm$^2$,
respectively.

The shrinkage of the distance to the horizon by the violent 
acceleration allows us to probe other aspects of gravitational 
physics. For example, Arkani-Hamed \etal~\cite{arkani00} suggested that 
extra dimensions of the quantum gravity may have manifestations in a 
relatively low extra dimension $(n)$ in this four-dimensional world. 
The distance over which this may be manifested is
\bq
r_{n}\sim 10^{30/n-17}\ \mbox{cm}.\label{ark1}
\eq
The distance to the horizon created by the intense laser acceleration 
of an electron
\bq
d={c^{2}\over a_{e}}={\lambda\over 2\pi}{1\over a_{0}},\label{ark2}
\eq
could exceed the above distance Eq.~(\ref{ark1}) if $n$ is less than or 
equal  to 4.

\bigskip

Zetta- and exawatt-lasers will allow us a glimpse into some of the 
most energetic and enigmatic phenomena of astrophysics such as GRB's, and their associated effects on EHECR, 
the final energy frontier in the Universe (`Extreme Universe' in the highest energy and from the cosmological
distance).  We marvel at such fascinating possibilities of exploring fundamental physics that are enabled by the
extreme high-field science developments by zetta and exawatt lasers, even though we have just barely scratched the
surface so far on this subject. It is also noted that this laser could bring many frontiers of contemporary physics,
i.e.~particle physics, nuclear physics, gravitational physics, nonlinear field theory, ultrahigh pressure physics,
relativistic plasma and atomic physics, astrophysics, and cosmology together. Because of such significant potential
scientific impacts, though it amounts to nontrivial efforts and developments, it seems worthy of further and more
serious consideration of this extreme high-field science.\\

We appreciated the comment by Prof.~R. Siemann on polarized positron 
sources and by Profs.~N. Fisch, M. Key, and P. Mulser on fast ignition.  This work was supported in part by the U.S. Dept.~of Energy
Contracts~DE-FG03-96ER-54346,  W-7405-Eng.~48, and in part by the National Science Foundation Grant No.~STC
PHY~8920108.

\clearpage

\newcounter{figlist}
\subsection*{FIGURE CAPTION}
\begin{list}%
{FIG.~\arabic{figlist}.}{\usecounter{figlist}
                \setlength{\labelwidth}{.55in}
                \setlength{\leftmargin}{.60in}}

\item Laser-focused intensity vs.~years for table-top systems.This
system shows the dramatic increase in intensities experienced over the
past few years. The figure shows also that the maximum intensity for 
table-top systems will be reached in few years. For the systems we propose, the
laser intensity will far exceed these values.

\end{list}


\clearpage

\baselineskip 22pt
\begin{thebibliography}{55}

\bibitem{perry94} D. Strickland and G. Mourou,  Opt. Comm {\bf 56}, 219 
(1985).

\bibitem{mourou98} G. A. Mourou, C. P. J. Barty, and M. D. Perry,  \emph{Physics Today}, January 1998.

\bibitem{dubeis92} A. Dubeis \emph{et al.}, Opt. Comm. {\bf88}, 437 (1992).

\bibitem{morgner99} U. Morgner, F. X. K\"atner, S. H. Cho, Y. Chen, A. H. Haus, J. G. Fujimoto, and E. P. Ippen,  Opt. Lett. {\bf24},
411 (1999).

\bibitem{druon98} F. Druon, G. Cheriaux, J. Faure, J. Nees, M. Nantel, A. Maksimchuk and G. Mourou,  Opt. Lett. {\bf23},
1043-1045 (1998); O. Albert, H. Wang, D. Liu, Z. Chang, and G. Mourou, Opt. Lett. {\bf125}, 1125 (2000).

\bibitem{ross97} I.N. Ross, \emph{et al.}, Opt. Comm. {\bf144}, 125 (1997).

\bibitem{du94} D. Du, X. Liu, G. Korn, J. Squier, and G. Mourou, Appl. Phys. Lett. {\bf64}, 3071 (1994).

\bibitem{stuart95} B. Stuart, M. Feit, A. Rubenckik, B. Shore, and M. Perry, Phys. Rev. Lett. 
{\bf74}, 2248 (1995).

\bibitem{migus} A. Migus, Private Communications.

\bibitem{felt} M. Felt, Private Communications.


\bibitem{bloem65} N. Bloembergen, \emph{Nonlinear Optics},  (Addison-Wesley, Reading, 1965).

\bibitem{tajima79} T. Tajima and J. M .Dawson, Phys. Rev. Lett. {\bf43}, 267 (1979).

\bibitem{sprangle87} P. Sprangle, C. M. Tang, and E. Esarey, IEEE Trans. 
Plasma Sci. {\bf15}, 145 (1987); D. C. Barnes, T. Kurki-Suonio, and T. 
Tajima, \emph{ibid.} {\bf15}, 154 (1987).

\bibitem{ashour81} M. Ashour-Abdalla, J. N. LeBoeuf, T. Tajima, J. M. Dawson, and C. F. Kennel,
Phys. Rev. A {\bf23}, 1906 (1981).

\bibitem{tajima85} T. Tajima, Laser Part. Beams {\bf3}, 351 (1985).

\bibitem{scheid89} W. Scheid and H. Hora, Laser Part. Beams {\bf7}, 315 (1989).

\bibitem{rau97} B. Rau,T. Tajima, and H. Hojo, Phys. Rev. Lett. {\bf78}, 3310 (1997).

\bibitem{lawson79} J.D. Lawson, IEEE Trans. Nucl. Sci. {\bf NS-26}, 4217 (1979); 
P.M. Woodward, J. IEE {\bf93}, Part III A, 1554 (1947).

\bibitem{sato72} H. Sato and T. Tati, Prog. Theor. Phys. {\bf47}, 1788 (1972).

\bibitem{nakajima01} K. Nakajima, this Proceedings (2001).

\bibitem{tabak94} M. Tabak, \emph{et al.}, Phys. Plasmas {\bf 1}, 1626 
(1994).

\bibitem{kodama01} R. Kodama, \emph{et al.}, Nature {\bf 412}, 798 (2001).

\bibitem{horton86} W. Horton and T. Tajima, Phys. Rev.~A {\bf 34}, 4110 
(1986).

\bibitem{key99} M. Key, \emph{et al.}, in \emph{First Int'l. Conf. Inertial Fusion Sci. Appl.} 
(Bordeaux, France 1999).

\bibitem{maksim00} A. Maksimchuk, \emph{et al.}, Phys. Rev. Lett. {\bf84}, 4108 (2000).

\bibitem{clark00} E. L. Clark, \emph{et al.}, Phys. Rev. Lett. {\bf85}, 1654 (2000).

\bibitem{snav00} R.A. Snavely, \emph{et al.}, Phys. Rev. Lett. (2000).

\bibitem{roth01} M. Roth, \emph{et al.}, to be published in \emph{Varenna Proceedings}, 
eds.`G. Mourou, T. Tajima, and M. Lontano (American Institute of Physics, New York, 2001).

\bibitem{esirk99} T. Zh. Esirkepov, T.V. Liseikina, F. Califano, N. M. Naumova, V. A. Vshirkov, F. Pegoraro, 
S. V. Bulanov,  JETP Lett. {\bf70}, 82 (1999); also A. Pukhov \emph{et al.} 
in Max-Planck, Y. Ueshima \emph{et al.}  in JAERI, S. Wilks
\emph{et al.} in LLNL, among other groups, have shown similar results.

\bibitem{koga} J. K. Koga \emph{et al.}, to be published in 
\emph{Varenna Proceedings, ibid}.

\bibitem{telnov82} I. Ginzburn, G. Kolkin, V. Serbe, and V. Telnov, 
JETP Lett. {\bf34}, 491 (1982).

\bibitem{coleman99} S. Coleman and S.L. Glashow, Phys. Rev.~D {\bf 59}, 
116008 (1999).

\bibitem{kish00} Y. Kishimoto and T. Tajima, in \emph{High Field Science}, eds.~T. Tajima, K. Mima, and 
H. Baldis, (Kluwer, NY, 2000) p.~83.

\bibitem{taguchi}T. Taguchi and K. Mima, \emph{ibid}., p.~163.

\bibitem{cowan00} T. E. Cowan \emph{et al.}, Phys. Rev. Lett. {\bf84}, 903 (2000).

\bibitem{udagawa96} T. Udagawa, private communication (1996).

\bibitem{bula96} C. Bula, \emph{et al.}, Phys. Rev. Lett. {\bf76}, 3116 (1996).

\bibitem{schwinger51} J. Schwinger, Phys. Rev. {\bf82}, 664 (1951).

\bibitem{unruh76} W. Unruh, Phys. Rev. D. {\bf14}, 870 (1976).

\bibitem{hawking74} S. W. Hawking, Nature (London) {\bf248}, 30 (1974).

\bibitem{chen99} P.S. Chen and T. Tajima, Phys. Rev. Lett. {\bf83}, 256 (1999).

\bibitem{arkani00} N. Arkani-Hamed, S. Dimopoulos, and G. Dvali, Phys. 
Rev. Lett. {\bf84}, 586 (2000).



\end{thebibliography}
\end{document}